\documentclass[conference]{IEEEtran}
 
\usepackage{url}
\usepackage{graphicx}
\usepackage{amsmath}
\usepackage{algorithm, algpseudocode}
\usepackage{verbatim}
\usepackage{tabularx}
\usepackage{array,multirow}
\usepackage{color}
\usepackage{comment}

\title{Making Honey Files Sweeter: SentryFS - A Service-Oriented Smart Ransomware Solution}

\author{
	\IEEEauthorblockN
	{
		Abdul Rahim Saleh\IEEEauthorrefmark{1},
		Gihad Al-Nemera\IEEEauthorrefmark{1},
		Saif Al-Otaibi\IEEEauthorrefmark{1},
		Rashid Tahir\IEEEauthorrefmark{1},
		Mohammed Alkhatib\IEEEauthorrefmark{1},
	}
	\IEEEauthorblockA
	{
		\IEEEauthorrefmark{1}University of Prince Mugrin, KSA, \emph{(3810162, 3810226, 3810200, r.tahir, m.al-khatib)@upm.edu.sa}
	}
	
	\vspace*{-7mm}
}

\definecolor{mygreen}{RGB}{53, 173, 45}

\begin{document}

\maketitle

\begin{abstract}

The spread of ransomware continues to cause devastation and is a major concern for the security community. An often-used technique against this threat is the use of honey (or canary) files, which serve as ``trip wires'' to detect ransomware in its early stages. However, in our analysis of ransomware samples from the wild, we discovered that attackers are well-aware of these traps, and newer variants use several evasive strategies to bypass traditional honey files. Hence, we present the design of SentryFS - a specialized file system that strategically ``sprays'' specially-crafted honey files across the file system. The canaries are generated using Natural Language Processing (NLP) and the content and the metadata is constantly updated to make the canaries appear more attractive for smarter ransomware that is selective in choosing victim files. Furthermore, to assist with the management of the honey files, SentryFS connects with an anti-ransomware web service to download the latest intelligence on novel ransomware strategies to update the canaries. Finally, as a contingency, SentryFS also leverages file clones to prevent processes from writing to files directly in the event a highly stealthy ransomware goes undetected. In this case, the ransomware encrypts the clones rather than the actual files, leaving users' data unmodified. An AI agent then assigns a suspicion score to the write activity so that users can approve/discard the changes. As an early-warning system, the proposed design might help mitigate the problem of ransomware.

\end{abstract}

\section{Introduction}


In recent years, ransomware has emerged as a major threat for organizations and enterprises. Lost data and crippled operations result in huge financial losses for the companies targeted. In fact, the attacks are so devastating that 1 in every 5 of the businesses affected by a ransomware ends up shutting down permanently~\cite{veriato}. Furthermore, the ``business model'' behind these attacks is so lucrative for the attackers that incidents have become rampant. For instance, according to a recent report (2020) from BlackFog security firm, there is a ransomware attack every 11 seconds in the wild~\cite{blackfog}. 
Clearly, the stakes are rising and the cyber security community needs to develop effective mechanisms to mitigate the rise of ransomware.


To this end, security vendors have developed various solutions. Popular among them are those that rely on redundantly backing up data to quickly recover from a disaster in the event of a compromise. Similarly, other vendors rely on early detection mechanisms coupled with some sort of rapid response system that brings the business operations back online before any serious losses are incurred~\cite{veriato}. One example of the latter is the use of honey files, which are dummy files deployed across the system to serve as trip wires. If a process updates one of these honey files, an alarm is raised to signal a suspicious event. Since no legitimate process has need of updating the honey files, it is highly likely that ransomware is at play. However, given that there is an uptrend in ransomware infections, traditional mechanisms, including honey files, need more improvements to stay up-to-date with evolving ransomware variants that are smarter.

\begin{figure*}[t]
	\centering
	\includegraphics[width=\linewidth]{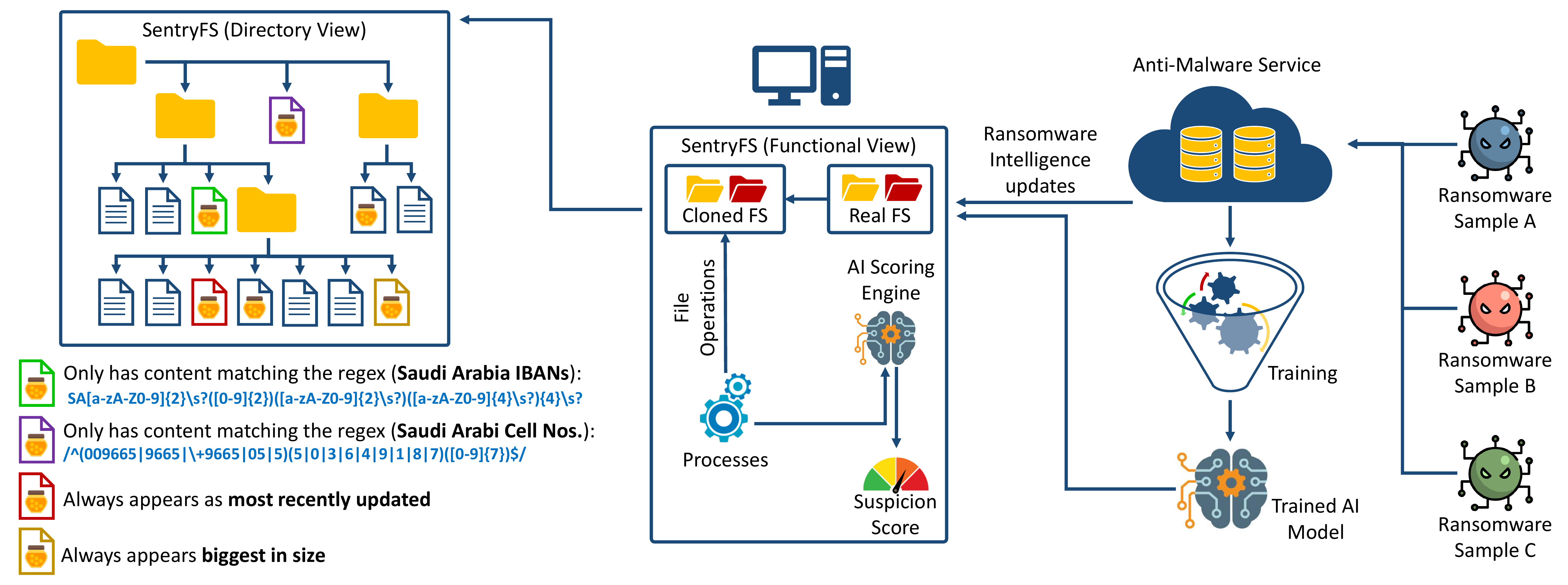}
	\vspace{-.25in}
	\caption{Overview of the Sentry File System (Functional View): Each SentryFS instance contains a set of real files and cloned files. Write operations are performed on the clones only. The anti-malware service shares updates on ransomware profiles and SentryFS makes the necessary updates to the honey files. Furthermore, each SentryFS instance also houses an AI scoring agent, which in turn contains a fully trained model shipped by the anti-ransomware service. Figure also shows the strategic placement of honey files (SentryFS - Directory View). The green canary contains artificially generated content that resembles IBANs of Saudi account holders. Similarly, the purple canary contains synthesized Saudi cell phone numbers.}
	\vspace{-.1in}
	\label{fig:attack}
\end{figure*}

Hence, in this paper, we analyze real-world ransomware samples to demystify the criteria used to select a victim file. We find that some smarter variants carefully select victims based on certain pre-defined criteria to avoid making ``noise''. 
Based on these findings, we propose our own specialized file system, SentryFS, which strategically creates and places honey files in every folder (or user-marked folders) so as to have a greater appeal towards smarter variants of ransomware. 
Our technique relies on using NLP and regexes to generate the content of the honey files and carefully manipulates the metadata to satisfy certain criteria. Furthermore, the honey files are constantly updated based on feedback received from an anti-ransomware service. The service analyzes real-world samples from the wild, digs out the file-access patterns and the selection criteria used to shortlist victim files and shares the intelligence with authenticated instances of SentryFS. Finally, as an extra layer of safeguard, SentryFS creates a virtual clone of a file anytime a process performs a write operation on the file. This clone is marked as the latest copy but never written back to the actual file system till the user approves the change. SentryFS helps the user make this decision by providing a suspicion score that attempts to capture the characteristics of the write operation in question. 


\section{Study of Ransomware \& Design of SentryFS}

\noindent\textbf{Analysis of Ransomware Samples:} To create effective honey files, we first need to understand how victim files are selected. To this end, we analyzed samples from around two-dozen ransomware families and found that a few smarter variants were examining the metadata (file size, date accessed, extension type, etc.) and the content (by matching against certain regexes or a word list) to choose which files to encrypt. This allowed the malware to bypass traditional canaries as they often do not match the metadata criteria and have garbage content. Going forward, we postulate that attackers could specify a criteria like the top \textit{\textbf{k}} files that are most recently modified in a folder, or files that have a certain type of content, such as IBAN numbers, which matches a fixed regular expression. All these tricks, would bypass conventional canary file mechanisms due to their static nature. Similarly, attackers can also select victim files by exploiting the automatic naming convention used by various devices, such as when a user copies data from a camera to their Windows machine. The camera will most likely have an algorithm to name the users' pictures automatically. In the same vein, certain apps have their own automatic naming conventions, which can be leveraged. These conventions can be easily embedded into the search criteria of the ransomware in the form of a regex or text matching ruleset. Clearly, these infection patterns and selection criteria can be leveraged to make honey files ``sweeter'' for the attacker and detect ransomware in its early stages.


\noindent\textbf{Processing and Management of Honey Files:} The creation, modification and placement of these honey files is of paramount importance as revealed by the ransomware analysis. SentryFS visits each user-defined directory in the file system that houses important user data, and creates honey files by following the popular naming conventions and patterns discussed previously (see the working in Figure~\ref{fig:attack}). It has its own text synthesizing agent to create content that matches certain regular expressions (such as cell phone numbers). For now, the synthesizing agent uses text from various blog posts and novels. However, once the work is completed, we will have an NLP text generator that will synthesize the content of the honey files. SentryFS can also continuously modify the metadata of honey files (updating the last modified date/time or changing the size) to make them more attractive to certain types of ransomware that select victim files based on the metadata. All these operations are performed by a software agent integrated into the file system.

\noindent\textbf{Anti-Ransomware Service:} Like other forms of anti-virus software, SentryFS connects to an online service that provides intelligence on how to prevent novel forms of ransomware from infecting the system. This service collects samples from the wild, analyzes them and determines what the victim selection criteria is, what the file access patterns are, are naming conventions targeted, should a regex be used or do we need to generate custom text using NLP, etc. All these findings are then shared with authenticated instances of SentryFS, which use the information to update the corresponding canaries. If SentryFS discovers a ransomware instance, the binary is promptly shared with anti-ransomware service for analysis. Furthermore, the service is also responsible for generating the AI-based scoring agent. The goal is to train a model that can examine the behavior of a process that has updated some files and give a score based on how suspicious the activity appears. To this end, we will train a classifier (either using ensemble learning methods or deep learning models with varying architectures) to output a probability score indicating whether the process is malicious or not. For training, we will use our own dataset, which was collected during analysis of the malware samples on air-gapped machines in our lab. 

\noindent\textbf{Virtual Clones:} In addition to honey files, SentryFS also leverages virtual clones of files~\cite{shield}. This prevents an evasive ransomware from directly encrypting the files of a user. Furthermore, a profile of the write operations of an evasive ransomware caught manually by the user, will then disseminated to other users via the anti-ransowmare service. 

\noindent\textbf{Conclusion and Future Work:} We perform a basic study to understand the infection patterns and victim-selection strategies of ransomware to improve honey files. The findings have been cataloged and baked into the prototype SentryFS system. This is still a work-in-progress but currently, the system can generate and deploy certain types of honey files and raise alerts if any of the honey files are updated. Going forward, we will be working on the anti-ransomware service and the NLP text synthesizing agent.




\bibliography{bibliography}

\begin{thebibliography}{1}

\bibitem{veriato}
``{Veriato RansomSafe}.'' \url{https://tinyurl.com/y33c6sa6}.

\bibitem{blackfog}
``{The State of Ransomware in 2020}.'' \url{https://tinyurl.com/yyu4q6h4}.

\bibitem{shield}
A.~Continella, A.~Guagnelli, G.~Zingaro, G.~De~Pasquale, A.~Barenghi,
  S.~Zanero, and F.~Maggi, ``Shieldfs: A self-healing, ransomware-aware
  filesystem,'' ACSAC, 2016.

\end{thebibliography}
\bibliographystyle{ieeetr}

\end{document}